\documentclass{article} 
\usepackage[english]{babel}
\newcommand{\lb}{\langle}
\newcommand{\rb}{\rangle}

\newcommand{\beq}{\begin{equation}}
\newcommand{\eeq}{\end{equation}}
\newcommand{\lbl}{\label}
\newcommand{\beqnar}{\begin{eqnarray}}
\newcommand{\eeqnar}{\end{eqnarray}}
\newcommand{\beqnars}{\begin{eqnarray*}}
\newcommand{\eeqnars}{\end{eqnarray*}}

\newcommand{\goesto}{\rightarrow}
\newcommand{\s}{\\[1ex]}
\newcommand{\re}[1]{(\ref{#1})} 
\newcommand{\q}{\quad}

\newcommand{\tnr}{\otimes}

\newcommand{\dajA}{\mbox{\boldmath $\hat{A}$}} 
\newcommand{\dajB}{\mbox{\boldmath $\hat{B}$}} 
\newcommand{\dajU}{\mbox{\boldmath $\hat{U}$}} 
\newcommand{\dajI}{\mbox{\boldmath $\hat{I}$}} 
\newcommand{\dajP}{\mbox{\boldmath $\hat{P}$}} 
\newcommand{\dajrho}{\mbox{\boldmath $\hat{\rho}$}} 
\newcommand{\dajE}{\mbox{\boldmath $\hat{E}$}}
\newcommand{\dajEj}{\mbox{\boldmath $\hat{E}_j$}}
\newcommand{\dajM}{\mbox{\boldmath $\hat{M}$}} 
\newcommand{\dajMj}{\mbox{\boldmath $\hat{M}_j$}}

\newcommand{\tr}{\mbox{Tr}}

\newcommand{\diag}{\mbox{\bf diag}}
\begin{document} 
\title{
{Proof gap in ``Sufficient conditions for uniqueness of the Weak
Value'' by J.\ Dressel and A.\ N.\ Jordan, J.\ Phys.\ A {\bf 45} (2012)
015304} 
\author{Stephen Parrott\thanks{For contact information, 
go to http://www.math.umb.edu/$\sim$sp}} 
} 
\date{February 17, 2013} 
\maketitle 
%
%
\begin{abstract}
The commented article attempts to prove a ``General theorem'' giving
 sufficient conditions under which a previously introduced
``general conditioned average'' ``converges uniquely 
to the quantum weak value in the minimal disturbance limit.''
The ``general conditioned average'' is obtained from a positive operator
valued measure (POVM) $\{\dajEj(g)\}^n_{j=1}$ depending on 
a small ``weakness'' parameter $g$.
We point out that unstated assumptions in  the presentation of the  
``sufficient conditions'' make them appear much more general than they
actually are. Indeed, the stated ``sufficient conditions'' strengthened by
these unstated assumptions seem very close
to an assumption that the POVM operators $\dajEj(g)$  be  
{\em linear} polynomials 
(i.e., of first order in $g$). 
Moreover, there appears to be a critical error or gap in the attempted proof, 
even assuming a linear POVM.  
A counterexample to the {\em proof} of the 
``General theorem'' (though not to its conclusion) 
is given.  Nevertheless, I conjecture that the conclusion is actually true for 
{\em linear} POVM's whose contextual values are chosen 
by the commented article's 
``pseudoinverse prescription''.
\end{abstract} 
\noindent
\section{Relation between traditional ``weak measurement'' theory and
the ``contextual value'' approach of \protect{\cite{DJ}} } 
\subsection{General introduction}
This is an expanded version of a paper submitted to 
J.\ Phys.\  A commenting on \cite{DJ} (called DJ below).   
DJ attempts to refute counterexamples given in \cite{parrott4} 
to a ``General theorem'' (GT).  The ``Comment'' paper discusses 
 the validity of 
these counterexamples and gives a new counterexample to the {\em proof}
(though not to the conclusion) of the GT.

The submitted paper had to be written more tersely in order
to keep its length appropriate for a ``Comment'' paper and may  
be incomprehensible to anyone not already familiar with DJ.  
This expanded version adds the present Section 1 introduction together with 
an appendix which analyzes in  
detail some logical problems with the stated hypotheses of the GT. 

For orientation I first give a brief description of my view of the notion 
of ``contextual values'', introduced in \cite{DAJ} (called DAJ
below)%
\footnote{ 
The reader should be warned that DAJ is vaguely written with many 
errors and omissions of important definitions and hypotheses.  
} 
and expounded by Dressel and Jordan in more detail 
in \cite{DJ2}%
\footnote{\cite{DJ2} is written in an unusual, complicated notation
different from both DAJ and DJ.  
It contains what the authors characterize
as  a ``slight generalization'' of the ``General theorem'' of the first 
arXiv version of DJ. The discussion of
the GT in \cite{DJ2} gives no indication that its validity is disputed 
in \cite{parrott4}. 
}
and the paper DJ under review \cite{DJ}.
My views are presented in more detail in  
\cite{parrott1}%
\footnote{
The reader should be warned that the six versions of \cite{parrott1} 
were written 
over a period of months as I tried to make sense of the vaguely 
written and error-ridden DAJ, and an evolution
of its ideas from earlier to later versions will be apparent.  The presentation
may seem unusual in that introductions to the later versions were simply
prepended to rewritten earlier versions.  The presentation of \cite{parrott4}
is more concise.  
}    
and \cite{parrott4}.  
The review assumes that the
reader has  some familiarity with the ideas of weak measurement.
A more leisurely exposition of these can be found in 
\cite{parrott5} and references cited there.
To minimize confusion, 
I try to use the notation of DJ wherever practical, 
even though it is not the notation that I would choose.  

One minor exception is that 
I usually write $\lb u| \dajA v \rb$ instead of  $\lb u | \dajA | v \rb$
as in DJ's (1.1).  The reason is that all the type was set before noticing
this small difference, and attempting to change it 
risks more confusion than retaining it in case some instances which 
should be changed go unnoticed. 
\subsection{Weak measurement}

``Weak measurement'', introduced in \cite{AAV},  is in part a technique  
for measuring the expectation of a quantum observable $\dajA$ 
in a given state $s$ without appreciably changing the state.  
This can be accomplished as follows.

Suppose the observable $\dajA$ operates on a Hilbert space $S$.  
Couple $S$ to an auxiliary ``meter space'' $M$, obtaining a new Hilbert
space $S \tnr M$ which is the tensor product of $S$ and $M$.   

With each state $s$ of $S$, associate a slightly entangled state
$\dajU s \in S \tnr M$, where $\dajU$ is a isometry%
\footnote{An {\em isometry} $\dajU$ is a linear transformation which 
preserves inner products: $\lb \dajU v | \dajU w \rb = 
\lb v | w \rb $ for all $v,w$.    The only difference between an
isometry and a unitary operator is that an isometry need not be 
surjective (i.e., ``onto'').
}
from $S$ to $S \tnr M$.
Find a ``meter observable'' $\dajB$ on $M$ such that 
the expectation of $\dajI \tnr \dajB$ in the state $\dajU s$ is almost the
same as the expectation of $\dajA$ in the state $s$, where $\dajI$ generically 
denotes the identity operator on whatever space is relevant in the context
(in this case $S$). 

To make this precise, 
introduce a small real ``weak measurement'' parameter
$g$ with $\dajU = \dajU(g)$ depending on $g$ .
In terms of this parameter, ``slightly
entangled'' is interpreted as 
\beq
\lbl{eq493}
\lim_{g \goesto 0} \dajU(g) s = s \tnr m \q,
\eeq
an unentangled product state, where $m$ is some state of $m$.  

Denote the projector onto the subspace spanned by a nonzero vector $v$
as $\dajP_v$. 
A routine calculation shows that the preceding paragraph
 guarantees that the (mixed) state of $S$
corresponding to $\dajU (g) s$, namely $\tr_M 
\dajP_{ \mbox{ \boldmath $\scriptstyle \hat{U}$} (g) s} $, 
where $\tr_M$ denotes partial trace with respect to $M$, 
approaches $\dajP_{s}$ 
(i.e., the original pure state 
$s$ written in mixed state notation) as $ g \goesto 0$.  
Similarly 
``the expectation of $\dajI \tnr \dajB$ in the state $\dajU(g) s$ is almost the
same as the expectation of $\dajA$ in the state $s$'' is interpreted as 
holding exactly in the limit $g \goesto 0$:
\beq
\lbl{eq494}
\lim_{g \goesto 0} \lb \dajU(g)s , (\dajI \tnr \dajB) \dajU(g)s \rb = 
\lb s | \dajA s \rb
\q.
\eeq	

The mathematics of DJ can easily be made rigorous only 
under the assumption that all Hilbert spaces
occurring are finite dimensional, so that all observables have discrete 
spectra.  
Let $\{ f_j \}^n_{j=1}$ denote a collection of orthonormal eigenvectors of 
$\dajB$, with $\alpha_j $ the corresponding eigenvalues, which we assume 
distinct%
\footnote{The slightly subtle reason is discussed in \cite{parrott1}.
}
for expositional simplicity.
We shall allow the 
$\alpha_j = \alpha_j(g)$ to depend on $g$, which implies that 
$\dajB = \dajB(g)$ also depends on $g$.  In principle, one could also allow
the eigenvectors
$f_j$ to depend on $g$, but for simplicity we assume that they are constant.  
(This assumption can be justified by making appropriate identifications.) 

Write 
\beq
\lbl{eq500}
\dajU(g) s = \sum_j \dajMj (g) s \tnr f_j
\q,
\eeq
where this defines the ``measurement'' operators $\dajMj(g)$ on $S$.
These measurement operators define a positive operator valued measure 
(POVM) $\{\dajEj(g)\}$ on $S$ by $\dajEj(g) := \dajMj^\dag(g) \dajMj(g)$.  
The probability $P(j)$ that a measurement of $\dajI \tnr \dajB$ in state 
$\dajU(g) s$  will produce result $j$ is the norm-squared of the 
$f_j$-component of \re{eq500}:
\beq
\lbl{eq505}
P(j) = 
|\dajM_j(g)s|^2 = \lb s | \dajEj(g) s \rb   
\q.
\eeq 
From \re{eq505}, the expectation 
$\lb \dajU(g)s , (\dajI \tnr \dajB) \dajU(g)s \rb $ 
of $\dajI \tnr \dajB$ in the state $\dajU(g) s$
is 
\beq
\lbl{eq520}
\lb \dajU(g)s , (\dajI \tnr \dajB) \dajU(g)s \rb = 
\sum_j \alpha_j P(j) = \sum_j \alpha_j(g) \lb s | \dajEj(g) s \rb
\q. 
\eeq 
Since the probabilities $P(j)$ depend only on data in $S$, 
by allowing the eigenvalues $\alpha_j = \alpha_j(g)$ to depend
on $g$, we might hope to choose them to satisfy the desired relation
\beq
\lbl{eq525}
\lb \dajU(g)s , (\dajI \tnr \dajB (g) ) \dajU(g)s \rb = 
\sum_j \alpha_j(g) P(j) = \lb s | \dajA s \rb \q,
\eeq 
which says that the expectation of the system observable $\dajA$
in the state $s$
could also be obtained by measuring the expectation of the meter observable
$\dajB (g)$ in the state $\dajU (g) s$.  The advantage of measuring  
$\dajB (g) $ instead of $\dajA$ is that for small $g$, the (unnormalized) 
state of $S$ after measurement result $j$ is obtained, namely 
$\dajM_j (g) s$, 
is very close to $s$ because of \re{eq493}.%
\footnote{This looks reasonable, but requires some calculation.} 
%

However, it is important to realize that only in special cases can one
hope to attain \re{eq525} for {\em all} small $g\neq 0$.  In general, 
given $\dajU (g)$, the required $\alpha_j (g)$ need not exist.  
The original formulation of weak measurement theory by Aharonov, 
Albert, and Vaidman  \cite{AAV} used
a particular $\dajU (g)$ originally proposed by von Neumann 
\cite{vN} which guarantees
\re{eq525}. 

Note, however, that if we are only interested in weak measurements,
(and in the opposite case, it is not clear why we would choose such a 
roundabout way to measure $\lb s | \dajA s \rb$), we do not need 
\re{eq525} for {\em all} $g$, but only for small $g\neq 0$.  That is, we
need only
\beq
\lbl{eq527}
\lim_{g \goesto 0} \lb \dajU(g)s , (\dajI \tnr \dajB (g) ) \dajU(g)s \rb = 
\sum_j \alpha_j(g) P(j) = \lb s | \dajA s \rb \q,
\eeq 
The above
formulation of the concept of ``weak measurement'' (which is similar to
\cite{parrott5}) requires only \re{eq527}. 

The difference between \re{eq525} and \re{eq527} might seem slight and 
technical, but it is actually substantial.  That is because it is much
easier to attain \re{eq527} than the stronger \re{eq525}, though given 
$\dajU (g)$, \re{eq527} is not guaranteed, either.

The above discussion sketches a formulation of weak measurement theory
which gives a more or less  direct translation into the language
of DAJ and DJ. 
Weak measurement theory is traditionally formulated in terms
of a ``system'' Hilbert space $S$ and a ``meter'' space $M$  with 
a ``meter observable'' $\dajB$.
The setup of DAJ and DJ replaces the meter space and meter observable
by a set of measurement operators on $S$, and
the strong equation \re{eq525} is {\em assumed}. 
The eigenvalues
$\alpha_j(g)$ of the meter observable $\dajB$ 
are renamed ``contextual values'' in the DAJ formulation.%
\footnote{The correspondence between projective measurements (identified 
with observables) in $S\tnr M$ and measurement operators in $S$ is
of course well known (cf. the text \cite{N/C}, section 2.2.8).  
Indeed, I suspect that it  may be 
what DJ is talking about in the second paragraph
of its section 2.  (It is hard to be sure because their symbol 
$\dajU_{sd}$ is not defined.) 
}
It seems clear that any statement about the meter observable
can be translated into a statement about measurement operators in $S$ 
and conversely,  

In summary, 
traditional measurement theory is formulated in a system+meter space
$S \tnr M$.  I view the contextual value theory of DAJ and DJ as a
conceptually similar formulation in system space $S$ alone.  Its 
chief merit (which should not be underestimated) seems to me to
be its simplicity.  
\subsection{Postselection}
Most applications of weak measurement theory involve more than mere 
weak measurement as described above.   Typically, after making a weak 
measurement one ``postselects'' to a given final state $s_f \in S$.
This means that one performs a second projective measurement
(with respect to the orthogonal decomposition $\{\dajP_{s_f}, 
\dajI - \dajP_{s_f} \}$) to see if after 
the first measurement, the system is in state $s_f$ (``success'') 
or a state orthogonal
to $s_f$ (``failure'').%
\footnote{For simplicity of exposition we restrict attention to postselection
to a pure state, as does DJ.  
DAJ considers postselection to a mixed state, 
but does not explain how this could be physically accomplished.} 
(It would take us too far afield to explain why one might want
to do this.)  This is done repeatedly starting with the same initial state
$s$, and only the results of ``successful'' trials 
are retained.  The (conditional)
expectation of the {\em meter} measurement given successful postselection 
is called
a ``weak value'' of the system observable $\dajA$.%
\footnote{The ``weak value'' is often confused with the 
conditional expectation of $\dajA$ (instead of the meter observable $\dajB$),
and it is important to keep the distinction in mind.
The conditional expectation of $\dajA$ must necessarily be a convex linear
combination of the possible values (eigenvalues) for $\dajA$, whereas 
the conditional expectation of $\dajB$ may lie far outside this set,
as the provocative title of \cite{AAV} suggests.
} 
This conditional expectation (with ``meter measurement'' replaced 
by ``measurement using given measurement operators'') 
is DAJ's ``general conditioned average'', 
which is routinely calculated.

Weak values are not unique; in general they depend on the measurement
procedure.  The seminal paper \cite{AAV} calculated (via questionable
mathematics) a particular weak
value for a particular measurement procedure.  Since this calculated 
weak value is generally nonreal (even though it is supposed to represent 
the procedure described above which would result in a weak value which
is manifestly real), most subsequent authors replace it with its real part.
The weak value calculated by \cite{AAV} is  equation (1.1) of DJ, 
written in a notation different from that
of DJ (and the present paper) which is common in the ``weak value''
literature:
$$
A_w = 
\frac{\lb \psi_f|\dajA|\psi_i\rb}
{\lb \psi_f | \psi_i \rb} 
\q. \eqno (1.1)
$$
Here $\psi_i$ represents the initial state of the system $S$ (called 
$s$ above), $\psi_f$ the postselected final state (called $s_f$ above),
and $A_w$ stands for ``weak value of $\dajA$''.  DJ's (1.2) is a generalization
of the real part of (1.1) to mixed states:
$$
_f \lb A \rb_w = 
\frac{\tr ( \dajE^{(2)}_f (\dajA \dajrho + \dajrho \dajA )}
{2 \tr ( \dajE^{(2)}_f \dajrho)}
\q.
\eqno{(1.2)}
$$
  
We shall refer to either the
real part of (1.1) or (1.2) as the ``traditional'' weak value (though I've not 
seen (1.2) in the literature prior to DAJ).  
Most of the ``weak value'' literature seems to implicitly assume 
that the traditional
weak value is the only possible weak value.

In the contextual value approach, it is natural to ask which
collections of  measurement operators $\{ \dajMj (g) \}$ will result 
in the traditional weak value in the limit $g \goesto 0$.  
DAJ claims without adequate proof that this will occur
when the measurement operators are positive.
DJ formulates and attempts to prove a ``General theorem'' (GT) 
with this conclusion. 
One might roughly summarize the GT by the statement that the 
traditional weak value is essentially inevitable when the measurement
operators are positive and commute with each other and the system observable
$\dajA$, and contextual values are chosen according to the
so-called ``pseudoinverse prescription''.

Counterexamples to the GT are given in  \cite{parrott4}. DJ attempts
to refute these counterexamples by reinterpreting (but unfortunately 
not restating in a logically precise way) 
the hypotheses of the GT given in the first preprint version of DJ,  
arXiv:1106.1871v1, to which \cite{parrott4} replied. 
The present work will make the reinterpretations explicit and give
a new counterexample to the {\em proof} of the GT (though not to 
its conclusion) under its reinterpreted hypotheses. 
\section{The ``Comment'' paper submitted to J.\ Phys.\ A} 
%
\subsection{Circumstances}

The following, from the next subsection \ref{subsecpaper} to the 
``Update'' just before the appendices, 
is essentially a ``Comment'' paper currently under review
by 
J.\ Phys.\ A (JPA), submitted on December 13, 2011. 
It is not identical because 
a few expositional improvements
have been made, but there are no differences of substance. 

Inquiries of  May 13 and July 18 
yielded  courteous responses, but no
substantive information about its  status except that it is 
being considered by a referee.  As of this writing, September 20, 2012, 
over nine months after initial submission, 
I have no idea what might be its status.

By comparison, DJ was accepted about a month after submission.  
It was first received on October 13, 2011, and 
I was notified on November 13 by J.\ Phys.\ A.\  
that it had been accepted.  

There is a story here, 
but this is not the place to tell it. 
Further information can be found on the ``papers'' page of my 
website, www.math.umb.edu/$\sim$sp.  
\\[2ex]
%
{\bf Note added February 8, 2013:}

The above were the circumstances when the previous Version 6 of this work was
posted.  Subsequently, the ``Comment'' was rejected, but not for any
substantive reason.  

In particular, there was no objection to 
its mathematics.  No referees' reports were included with the 
rejection, and the only ``reason'' given was that 
rejection was the ``most satisfactory'' course for the journal.  
After I specifically requested the referees' reports,  
JPA admitted that they had been unable to 
obtain any. Details, including the rejection letter, can be 
found at www.math.umb.edu/$\sim$sp. 

The authors (Dressel and Jordan) subsequently published in JPA
a ``Corrigendum'' \cite{corrigendum} which acknowledges a gap 
in the proof of the
``General theorem'' (GT) of DJ \cite{DJ}.  They present a Lemma 
which they think fills the gap, presented in a way which suggests
that it answers Version 6's  objections to DJ's attempted proof of 
its GT.  

However, I dispute that the Lemma does fill the gap in the proof of the GT.  
The authors have chosen to ignore Version 6's objection to DJ's attempted
proof.   

In the context of DJ's attempted proof, 
the Corrigendum's Lemma is essentially equivalent 
to the Conjecture which is the
centerpiece of the present work, though the statements of the
Lemma and Conjecture are superficially different.  
The Corrigendum does not mention the Conjecture,%
\footnote{ 
The Conjecture is prior to any public (and, to my knowlege, private) 
statement of DJ's Lemma. DJ learned of the Conjecture from privileged
information---a copy of the submitted ``Comment'' furnished them by JPA
before it was publicly posted.}
which Version 6 states that I believed I had proved.  

I believe that the Lemma of the Corrigendum, if correct, 
would only establish 
the GT for the
case of a ``linear'' POVM (defined  in subsequent subsections).  The
Conjecture stated that the GT would be true under this additional 
``linearity'' hypothesis.  

However, the proof of the Lemma appears to be fatally flawed due
to a calculational error.  This Version 7 adds a final Appendix 3
which pinpoints the error and gives a simple algebraic proof
of an important special case of the Lemma/Conjecture.%
\footnote{When Version 6
was posted, I had convinced myself that  
I could prove the full Lemma/Conjecture, but refrained from claiming this 
because I've not wanted to invest the time to write out the proof
in full.  It was not much trouble to write out the special case. 
}

I apologize to the reader for what may
seem like confusing updates like this one added here and there.  
I have put an enormous amount of work into trying to unravel the 
densely written claims of DJ, and I have neither 
the energy nor desire to rewrite the whole thing.  Also, I think
that the semi-chronological presentation may provide useful insights
into the procedures and professional standards of JPA.

Quite a few morals could be drawn from the history of the present work,
but I leave their formulation to the reader.  
To preserve the historical record,
I have not altered Version 6 from here to the final Appendix 3.
The only typo that I have noticed in Version 6 is one phrase 
``take into count'' where ``take into account'' was intended.  

\subsection{The submitted ``Comment'' paper (with minor corrections)}
\lbl{subsecpaper}
%
\subsubsection{Introduction to the submitted ``Comment'' paper}

Notation will be the same as in the
article under review \cite{DJ}, called DJ below.  To compress this Comment
to a traditional length, we must assume that the reader is already 
familiar with DJ.  
Its main purpose seems to be to justify a statement of \cite{DAJ} 
(called DAJ below)
that a ``general conditioned average'' introduced in DAJ 
``converges uniquely to the quantum weak value in the minimal disturbance
limit''. 
DJ formulates ``sufficient conditions'' as hypotheses for  
a ``General theorem'' (GT) with this statement as its conclusion.

For simplicity, we shall only consider the special case of 
DAJ and DJ's ``minimal disturbance'' condition
for which all measurement operators $\{\dajMj\}$ are positive. 
(All statements will also hold for DJ's slightly 
more general definition.) 
The associated positive operator valued measure 
(POVM) is $\{\dajEj \}$ with $\dajEj := \dajM^\dag_j \dajMj$.    
The measurement operators $\dajMj = \dajMj(g)$ depend on a small
``weakness'' parameter $g$ which quantifies the degree to which
the measurement affects the system being measured.  
Our ``minimal disturbance limit'' will refer 
to the so-called ``weak limit''
$g \goesto 0$ for positive measurement operators.

DAJ claims that under these assumptions, its 
``general conditioned average'' (corresponding to
what is more usually called a ``weak'' measurement followed by a postselection)
is given by the traditional ``quantum weak value'' 
(the real part of DJ's (1.1)) in the weak limit $g \goesto 0$:%
\footnote{The term ``minimally disturbing measurement'' for a 
measurement with positive measurement operators was used (and perhaps
coined) in the
recent book \cite{wiseman} of Wiseman and Milburn.  This
reference was unfortunately not cited in DAJ, which uses the 
term ``minimal disturbance limit'' without definition or intuitive 
explanation.  I've not seen the term used elsewhere in the literature
outside of DAJ and subsequent papers by its authors.  The technical
definition of the phrase ``minimally disturbing measurement'' as
referring to a measurement with 
{\em positive} measurement operators'' does not correspond to the
meaning which one might assume from ordinary usage of the words
``minimally disturbing''.  (This is discussed in more detail in
\cite{parrott1}, Section 11.) 
\s
For simplicity of exposition,
our definition of  ``minimal disturbance limit'' will be essentially
that of Wiseman and Milburn:  
the limit $g \goesto 0$ for {\em positive} measurement
operators, even though 
DJ uses a slightly more general definition.
All statements will also hold for DJ's definition.}
\begin{quote}
``This technique leads to a natural definition of a general 
conditioned average that converges uniquely to the quantum 
weak value in the minimal disturbance limit.''
\end{quote}

Counterexamples to this claim were given in 
\cite{parrott4}, examples which DJ attempts to refute by reinterpreting 
the hypotheses of its ``General theorem''(GT) given in the first
version of DJ, arXiv:1106.1871v1.  
Unfortunately, DJ does not make explicit this reinterpretation,  
but some such reinterpretation is necessary 
for their objection to make sense.

The mathematics of these counterexamples is undisputed;%
\footnote{However, DJ does correctly note a typo in the definition 
of one of the measurement operators in \cite{parrott4}; a $\sqrt{1/3}$
had been mistakenly written as $1/3$.  However the correct value
was used in the subsequent calculations, so apart from this single substitution,
no other alterations in the
argument of \cite{parrott4} are necessary. 
I thank the authors for this helpful correction.}
the only issue is
whether they satisfy the hypotheses of DAJ or DJ.  DJ correctly notes that 
the first counterexample using $2 \times 2$ measurement matrices 
does not satisfy what 
they call the ``pseudoinverse prescription'', but DAJ does not clearly state
this prescription as a hypothesis.%
\footnote{This is discussed in more detail in Appendix 2 and \cite{parrott4}.
}
The second counterexample using $3 \times
3$ matrices does satisfy the pseudoinverse prescription, so the following will 
deal exclusively with this counterexample.  

Contrary to claims of DJ, 
this counterexample  {\em is} valid when the hypotheses 
of the ``General theorem'' (GT) are interpreted {\em as written}, according to 
standard usage of logical language.  However, DJ's attempted refutation
of the counterexamples requires a great strengthening of one of these 
hypotheses, a strengthening not noted in DJ.  
We shall see that when so strengthened, the hypotheses of the GT 
seem very close to the assumption 
that the POVM must be a {\em linear} polynomial  
in the weak measurement parameter $g$, i.e., 
\beq
\lbl{eq65}
\dajEj(g) = \dajE^{(0)}_j + g\dajE^{(1)}_j \mbox{where 
$\dajE^{0}_j$ and $\dajE^{(1)}$ are constant operators.}
\eeq

The analysis leading to this conclusion will be straightforward 
and simple.  DJ's attempted proof of the GT is densely written, and  
our analysis of it must be correspondingly technical.
Although probably few readers will be 
sufficiently familiar with the proof to convince themselves
either of its truth or of the claim that there is a major error, 
I hope that the analysis may motivate anyone tempted to employ 
(or cite without comment) the 
``General theorem'' to first  
carefully scrutinize its proof. 
\subsubsection{Unstated hypotheses for the ``General theorem''}
\lbl{sec3}
The hypotheses of the ``General theorem''(GT) which will concern us are: 
\begin{quote}
\begin{description}
\item[\rm ``(iii)]  The equality $\dajA = \sum_j \alpha_j (g) \dajEj (g)$
must be satisfied, where the contextual values $\alpha_j (g)$ are 
selected according to the pseudo-inverse prescription.
\item[\rm (iv)] The minimum nonzero order in $g$ for all $\dajEj (g)$ 
is $g^n$ such that (iii) is satisfied.''
\end{description}
\end{quote}
``Minimum nonzero order'' is not a standard mathematical phrase, but
I take its occurrence in (iv) to mean that 
\beq
\lbl{eq10}
\dajEj(g) = \dajE_j^{(0)} + g^n \sum_{k=0}^\infty \dajE_j^{(k+n)}g^k 
\eeq
with $\dajE^{(m)}_j$ constant operators and $\dajE_j^{(n)} \neq 0$.  
This is the way the phrase is used (just after DJ's equation (5.2))
 in the attempted proof of the GT. 
It is used in exactly the same way on p.\ 11 of DJ's analysis 
of the counterexample,
(just before equation (7.4)) to conclude that the ``lowest nonzero order''
$n$
 for the counterexample is $n=1$.  This is important because the only way
to interpret (iv) as referring to truncations seems to require a completely
different interpretation of ``minimum nonzero order'', as will be discussed
more fully in  Appendix 1.  

Then the logical content of (iv) is that all $\dajEj$ have the {\em same}
minimum nonzero order, which is to be denoted $n$.%
\footnote{The restrictive phrase ``such that (iii) is satisfied'' 
is logically redundant,
since (iii) has {\em already} been assumed.  If the authors mean that
some alteration of (iii) is to be assumed, such as (iii) with the $\dajEj$
replaced by their truncations to order $n$ or  (iii) 
with the original contextual values
previously denoted $\alpha_j(g)$ replaced by others or some 
combination of these, then standard 
logical language requires that this be explicitly stated.  I have considered
several alternative interpretations of (iv), 
but all have led to inconsistencies with other parts of DJ.  In the absence
of requested clarification from the authors, I selected the one which seems 
most nearly consistent with the rest of DJ.}  
This is a strange and quite restrictive assumption for a ``General theorem'', 
but it will not be our main concern. 

When the hypotheses of the GT are given the standard
logical interpretation just described, the counterexample using $3\times 3$
matrices which DJ attempts to refute 
is indeed a counterexample to the GT. 
However, DJ's attempt to refute the counterexample appears to 
assume something like the following,
\begin{quote}
{\small Denote by $\dajE_j^\prime (g)$ the truncation of the series \re{eq10}
to its minimum nonzero order $n$, namely, 
\beq
\lbl{eq13}
\dajE^\prime_j (g) := 
\dajE_j^{(0)} + \dajE_j^{(n)}g^n \q.
\eeq
Then (iv) assumes (iii) with the $\dajEj(g)$ in (iii) replaced by
$\dajE_j^\prime (g)$, but with the contextual values $\alpha_j(g)$
unchanged (i.e., the contextual values for the truncated 
POVM $\{\dajE^\prime_j(g)\}$ are the same as for the original POVM
$\{ \dajEj(g) \}$).  More explicitly, it assumes that
\beq
\lbl{eq15} 
\sum_j \alpha_j (g) \dajE^\prime_j(g) = \dajA,  
\eeq 
where the $\alpha_j (g)$ satisfy the pseudo-inverse 
prescription for the truncated POVM.
}
\end{quote}

DJ's objection to the counterexample, given after its equation (7.4), 
is that it does not satisfy \re{eq15}. 
DJ does not explicitly say that the contextual values for the truncated
POVM are the same as for the original POVM, but that seems suggested by the 
fact that it uses the same symbols, $\alpha_j (g)$ for both.
Also, the details of DJ's attempted proof support
that interpretation.  

Next recall that (iii) assumes that the contextual values 
$\vec{\alpha} (g) = (\alpha_1(g), \ldots , \alpha_n(g))$ 
satisfy the ``pseudoinverse prescription'' 
\beq
\lbl{eq20}
\vec{\alpha} = F^+ \vec{a}, 
\eeq
where $\vec{a}$ is a list of eigenvalues for the system observable
$\dajA$, and $F^+$ is the Moore-Penrose pseudoinverse for 
the matrix 
\beq
\lbl{eq30}
F  = F(g) := [\vec{E}_1 (g), \ldots , \vec{E}_n (g)].
\eeq
Here the column vector $\vec{E}_j (g)$ is the list of eigenvalues for 
$\dajEj (g)$, and $F$ is the matrix composed of those columns.
Note that $F$ is $g$-dependent, but we write $F = F(g)$ only when necessary
to emphasize this point, to avoid possible confusion with the result of 
applying the matrix $F$ to a vector. 
If contextual values exist (in general, they don't), they are {\em
uniquely determined} by the ``pseudoinverse prescription'' \re{eq20}. 

Since the contextual values for the truncated POVM 
$\{ \dajE^\prime_j(g) \}$
are assumed the same as those for the original and to also satisfy the
pseudo-inverse prescription for the truncated POVM,  
we also have 
\beq
\lbl{eq40}
\vec{\alpha} = {F^\prime}^+ \vec{a}
\eeq
with 
\beq 
\lbl{eq50}
F^\prime := [\vec{E}^\prime_1 (g) , \ldots , \vec{E}^\prime_n (g)],
\eeq
where the $\vec{E}^\prime_j (g)$ are the column vectors of eigenvalues for 
$\dajE^\prime_j (g)$.  Equation \re{eq40} also {\em uniquely determines}
the contextual values $\alpha_j (g)$, so it would be surprising if  
{\em both} \re{eq20} and \re{eq40} would hold except in the trivial case
in which $\dajEj = \dajE^\prime_j$ for all $j$.  In that case, we can
make $\{\dajEj\}$ linear (i.e., of form (1)) by replacing the parameter
$g$ by a new parameter $h := g^n$, 
so for brevity we shall refer to this as the ``linear case''.  
The hypothesis that both do hold seems very close to a hypothesis 
that the original POVM be linear.  
Indeed, I do not know of any example of a nonlinear POVM
for which both \re{eq20} and \re{eq40} can hold.
\subsubsection{Error or gap in proof}
\lbl{sec4}
Readers thinking of  building on the work of DAJ and DJ may need to convince
themselves of the validity of its ``General theorem'' (GT).  
Since its attempted
proof is densely written, it may help to pinpoint what I think is a
critical error (or at least a serious gap), 
even under the strong hypothesis that the POVM is linear.  

This hypothesis is equivalent to the assumption that 
the matrix $F = F(g)$ determining the contextual values $\vec{\alpha}$
is first order in  $g$, 
in which case the minimum nonzero order
of $F$ which the proof calls $n$ is $n=1$.
To expose the gap, we use these
assumptions to rewrite the questionable part of the proof 
in a simplified form.  It applies to a matrix $F$ with singular value
decomposition $F = U \Sigma V^T$, where $\Sigma$ is a diagonal matrix
and ``$U$ and $V $ are orthogonal matrices''.   All of these matrices
depend on the weak limit parameter $g$. 

The contextual values $\vec{\alpha}$ (which the proof renames
$\vec{\alpha}_0$)  are determined by the pseudoinverse
prescription $\vec{\alpha} = \vec{\alpha}_0 = F^+ \vec{a}$, 
where $\vec{a}$ is the vector of eigenvalues
of $A$. Here $F^+$ is the Moore-Penrose pseudoinverse of $F$, given by
$F^+ = V\Sigma^+ U^T$, where $\Sigma^+$ is the diagonal matrix obtained 
from $\Sigma$ by inverting all its nonzero elements. 

In reading the following, please keep in mind that if correct, it should
apply to {\em any} matrix function $F = F(g)$.  
Although an $F = F(g)$ derived from a POVM has a 
special form given in part by DJ's preceding
equation (5.9), {\em nothing in the following proof fragment
uses this special form}.  

The proof mentions ``relevant'' singular values,
but for brevity I have omitted the definition of ``relevant'' 
(which does not involve the special form of $F$) 
because for the simple counterexample to be given, 
all we have to know is that a ``relevant'' singular value 
is a particular kind of singular value, as the syntax implies.
The simplified proof fragment is: 
\begin{quote}
{\small
Since the orthogonal matrices $U$ and $V$ have nonzero orthogonal limits
$\lim_{g \goesto 0} U = U_0$ and  
$\lim_{g \goesto 0} V = V_0$ , such that $U^T_0 U_0 = V_0V^T_0 = 1$, and
since $\vec{a}$ is $g$-independent, then the only poles in the solution 
$\vec{\alpha}_0 =  F^+ \vec{a} = V \Sigma^+ U^T \vec{a}$ must come from 
the inverses of the relevant singular values in $\Sigma^+$.

Therefore, to have a pole of order higher than $1/g^1$, 
there must be at least one relevant singular value with a leading order greater
than $g^1$.   
\\[1ex]
[This much seems all right, though many details are omitted, 
but I cannot follow
the next and last paragraph of DJ's attempted proof.] 
\\[1ex]
However, if that were the case then the expansion of $F$ to order $g^1$
would have a relevant singular value of zero and therefore could
not satisfy (5.12), contradicting the assumption (iv) 
about the minimum nonzero
order of the POVM.  Therefore, the pseudoinverse solution 
$\vec{\alpha}_0 = F^+ \vec{a}$ can have no pole with order higher
than $O(1/g)$ and the theorem is proved. 
} 
\end{quote} 
If correct, the above proof fragment would imply that 
if a linear matrix function
$F(g) = P + gQ $, with $P$ and $Q$ constant matrices, 
has a singular value with a leading order greater than $g^1$, then
it also has a singular value which is identically zero.  
(Put differently, the proof claims that 
if no singular value is identically zero for all $g$, 
then all singular values are $O(g^1)$.) 
However, it is easy to construct counterexamples such as  
\beq
\lbl{eq70}
F := 
\left[
\begin{array}{cc}
1+ g & 1 \\
-1   & -1+g  
\end{array}
\right], 
\eeq
which has singular values 
$[g^2 + 2  - 2 \sqrt{g^2 + 1}]^{1/2} =
 g^2/2 + O(g^4)$ and
$[g^2 + 2  + 2 \sqrt{g^2 + 1}]^{1/2} = 2 + g^2/2 + O(g^4)$.

Without performing the somewhat messy calculation of the singular values,
one can see directly from Cramer's rule that since $\det F(g) = g^2$,
$F(g)^{-1} \sim g^{-2}$ which 
would make the contextual values $\vec{\alpha} = F^{-1} \vec{a}$ asymptotic 
to $g^{-2}$.
The essence of the full proof of the GT is to show (continuing to 
assume $n=1$ for simplicity) that the 
contextual values are $O(1/g)$, which implies that the ``numerator correction''
of DJ's (5.7) vanishes in the limit $g \goesto 0$. 

Let us try to follow in detail the last paragraph of the proof in the 
context of the counterexample.  
Applied to the $F$ of \re{eq70}, the last paragraph asserts
that if $F$ has a singular value of leading order 
greater than $g^1$ (which it does), then ``the expansion of $F$ to order
$g^1$ would have a relevant singular value of zero $\ldots$''.  However,
this is wrong because 
the expansion of $F$ to order $g^1$ is $F$ itself, and all
singular values are positive for small $g \neq 0$.  

I suspect that the  last paragraph of the attempted proof 
may be based on an erroneous implicit assumption that truncating a 
$\Sigma$
corresponding to $F(g)$ will produce the $\Sigma$ for the truncated $F$, 
i.e., that truncation commutes with taking of singular values. 
Otherwise, how could one possibly relate the $\Sigma$ corresponding
to $F$ in (5.12) to the different $\Sigma$ corresponding to the linear
truncation of $F$? 
(Even assuming this relation, additional argument seems required to
justify  the last paragraph of the proof fragment.) 

To make the above more explicit, write $\Sigma = \Sigma(F)$ to indicate the 
dependence of the matrix $\Sigma$ of singular values on $F$, and write
$\tau(F)$ for the linear truncation of $F$. 
I can begin to make sense of the last paragraph 
only by assuming that  
$$
\Sigma(\tau(F)) = \tau(\Sigma(F));
$$ 
which says that the singular values for the truncated $F$ are the truncations
of the singular values for $F$.
The counterexample shows that this is false for its $F$ which satisfies  
$\tau(F) = F$: 
$$
\Sigma (\tau(F)) = \Sigma(F) =  
\left[
\begin{array}{cc}
2 + g^2/2 + O(g^4) & 0 \\
0 & g^2/2 + O(g^4)
\end{array} 
\right] 
\neq 
\left[
\begin{array}{cc}
2 & 0  \\
0 & 0 
\end{array} 
\right] = \tau(\Sigma(F)). 
$$

Recall that \re{eq15} was my best guess at the intended expansion of DJ's
hypothesis (iv) from its logical meaning.  
(A direct request to the authors to confirm or correct this was ignored.)
My next best guess would be that the $\alpha_j (g)$ in \re{eq15} might
represent contextual values for the truncated POVM $\{\dajE^\prime_j\}$ 
that would not necessarily be contextual values for the original POVM
$\{\dajEj\}$.  However, the above objection to the proof would still apply. 
\subsection{Further remarks on the relation of hypothesis (iv) to
a hypothesis that the POVM be linear}

Whatever the intended meaning of DJ's (iv),  in view of DJ's objection
to the counterexample, (iv) presumably 
imposes some condition on the linear truncation (still taking $n=1$ for
simplicity) of the original POVM $\{\dajEj(g)\}$. This seems an unreasonable
hypothesis for a theorem billed as ``General''.  Certainly the original
claim of DAJ that its ``general conditioned average'' ``converges uniquely
to the quantum weak value in the minimal disturbance limit'' gives no 
hint that unstated hypotheses necessary to validate the claim would 
fail to apply to simple cases such as the later counterexample 
of \cite{parrott4} with POVM which is
quadratic in $g$. 

DAJ gives the strong impression 
that the traditional weak value is essentially inevitable 
when the measurement operators are positive.  DJ gives the same impression
under the
newly added hypothesis that the measurement operators commute 
with each other and the system observable $\dajA$, along with explicit
assumption of  
the pseudoinverse prescription.  A main point of both
\cite{parrott4} and the present Comment is to dispel any such false 
impressions.  

There is nothing in DAJ even close to  hypothesis (iv) 
of DJ's ``General theorem''.  The precise meaning which the authors
attribute to (iv) is unclear, but if it is \re{eq15} or anything close to  
that, its physical meaning seems very obscure.  Given a POVM $\{\dajEj (g)\}$, 
contextual values need not exist.  
DAJ's ideas apply only to cases when they do exist.
The physical meaning of this is clear---only in that case will 
the average value $\tr [A\rho]$ of the system observable in any state 
$\rho$ equal the sum  $\sum_j \alpha_j \tr [\dajEj (g) \rho]$ 
of average POVM measurements weighted by the contextual values $\alpha_j$.   
But the physical meaning of {\em additionally} assuming that contextual values 
exist for a {\em truncation} of $\{ \dajEj (g) \} $  seems very obscure, 
and that appears to be an essential part of the authors' interpretation
of hypothesis (iv).

Another way to look at this is that (iv) seems to assume that if 
contextual values exist for a given POVM, then contextual values also 
exist for a linear POVM.  Viewed in this way, for practical purposes,
(iv) seems very close (though not precisely mathematically equivalent)
to a hypothesis that the POVM be linear.  
\subsection{Status of the ``General theorem'' for linear POVM's}
It should be emphasized that \re{eq70}  
is only a counterexample to DJ's attempted {\em proof}, 
not a counterexample to the {\em conclusion} of the GT under the assumption 
that the POVM is linear, i.e.,  of the form  \re{eq65}.
For a counterexample to the conclusion, one would need an $F$ which is
derived from a POVM. 

Actually, I conjecture that the conclusion that the 
``general conditioned average''
is given by DJ's (1.2) (i.e., the traditional weak value generalized
to mixed states) 
is {\em true} for {\em linear} POVM's under the
other hypotheses of DJ.
If so, its proof will surely have to use in some essential way
the special form of an $F$ which comes from a POVM 
(e.g., all rows sum to 1).  

I have sketched such a proof but have not
written it in detail, so I make no claims.  I will be happy to share 
the ideas of the proof with any qualified person who might be interested
in expanding on them.  They are not difficult, but
annoyingly detailed.  If I decide not to write them 
up in journal-ready detail myself, 
I may put a sketch of a proof on my website, 
www.math.umb.edu/$\sim$sp. 

A main aim of this Comment is to focus attention on the case of 
linear POVM's.  
If the conjecture is true, it might help to explain why (to my knowledge)
actual experiments have only observed the traditional weak value 
$ \Re ( \lb \psi_f| \dajA | \psi_i \rb / \lb \psi_f| \psi_i \rb)$, despite 
the fact that arbitrary weak values can be obtained from different
measurement procedures, as stressed by DJ.%
\footnote{See \protect{\cite{parrott5}} or the list of references [10] of DJ. 
} 
Since these 
experiments are difficult and have only recently been performed, 
perhaps they correspond
to the simplest POVM's, 
e.g., {\em linear} POVM's arising from positive measurement operators.  
\\[2ex]
\noindent
{\bf Update September 21, 2012:}

The arXiv version 1 of Dressel and Jordan's \cite{DJ2}, 
posted on October 3, 2011,
contains a ``Theorem'' which the authors characterize as a ``slight 
generalization'' of the ``General theorm'' (GT) of DJ \cite{DJ}, 
with essentially the same attempted proof 
that the present work criticizes.
There is no mention that this ``Theorem'' was disputed in
\cite{parrott4}, which is not even referenced.
(The omission was deliberate because the authors did not correct 
it in a later version after I pointed it out to them.)

Version 2 of \cite{DJ2}, posted in the arXiv on February 27, 2012 and 
published in Phys.\ Rev.\ A on Feb. 29, 2012, 
contains essentially the same ``Theorem''
but with a substantially different attempted proof.   Hypothesis (iv) 
has been somewhat clarified, and a lemma (their Lemma 1) has been added
which in the context of the proof is essentially the ``Theorem'' 
with a hypothesis that the POVM be 
linear, i.e., Lemma 1 is essentially equivalent to the Conjecture which is 
the centerpiece of this article.   

The Conjecture is not mentioned.  This can perhaps be understood from 
the fact that the authors learned of the present work and the Conjecture
from a review copy of the present work furnished them for their reply 
by JPA between
December 13, 2011 (when the present work was submitted and February 14, 2012
 (when the authors replied). According to published ethical standards of 
the American Physical Society, it is unethical to use privileged information
like review copies for professional advancement without the permission of 
the author of the reviewed work.  (My permission was never requested.) 
Or, perhaps the authors independently noticed that the proof of the GT 
in DJ \cite{DJ} was wrong and therefore thought 
it unnecessary to mention the Conjecture. 

Modulo technical mathematical details which I have not checked,
it looks as if the proof of their Lemma 1 could be correct.  If so, that 
will settle the Conjecture.
 
However, I think that the objection above to the proof of DJ's GT
also applies to the attempted proof of the full ``Theorem'' 
(i.e., for a not necessarily linear
POVM) of the published version of \cite{DJ2}.  That is, I don't agree
that the ``Theorem'' of the final version of \cite{DJ2} 
(nor the GT) has been proved.
However, that seems of minor consequence 
since hypothesis (iv) is so close
to an assumption that the POVM be linear. 

I have requested a copy of the authors'  reply to the review copy of the
present work, both from JPA and the authors.
JPA refused.  I get the impression that is  because they think it would
be unfair to furnish the ``Reply'' to a ``Comment'' before the ``Comment'' 
is accepted; otherwise the author of the ``Comment'' could modify it 
in light of the ``Reply''.  This would make
sense to me only if one adopted the attitude that a search 
for the truth of the ``General theorem'' (GT) must necessarily be adversarial.  
If the authors
have some valid objection to my claim that there is a serious gap in DJ's 
attempted proof 
of the GT, then, depending on the nature of the objection, either 
the present ``Proof Gap'' should be withdrawn, or it should be modified 
to take into count the objection (with proper acknowledgement of the authors' 
contribution).  It would be much more useful to readers that way, with
at least some agreement among all parties acknowledged. 

The authors have ignored my request for a copy.
The point is that I have made every effort to assure the accuracy 
of the present work. If I have fallen short in any way, it is 
not for lack of inquiry.

Readers who want to know whether the GT is sound without 
dissecting its dense and not entirely clearly written attempted proof will  
look to see if this ``Comment'' is ever published by JPA.  But it seems
quite possible that it will never be published, not because of any 
demonstrable error, but because no referee will be willing to 
invest the time to read the proofs in detail.
The fact that over nine months have passed
since the present work was submitted almost certainly indicates that
it has been, and remains, gathering dust on the desk of some referee
or editor of JPA.  It will probably remain that way indefinitely until someone
devises a pretext to accept or reject it without reading it.

Thus publication status is unlikely to be a reliable indication of 
correctness.  
A more reliable indicator may be
that the published version of \cite{DJ2} substantially 
alters DJ's attempted proof of the GT in ways which appear to 
respond to criticisms of the present work. 
That seems a tacit admission
that the present claim of a gap in the proof of the GT
has substance.

\section{Appendix 1: Guesses at the meaning of hypothesis (iv)} 

When I saw the grounds on which DJ disputed the counterexample
of \cite{parrott4}, I was stunned.
Never had I even considered the possibility that hypothesis (iv)
might refer to truncations, and had I considered it, I would have 
rejected it as implausible.  I still find it hard to imagine 
that {\em any} careful reader could confidently assert that 
(iv) referred to truncations, much less be confident of any 
definite meaning regarding truncations.   

After DJ was accepted and I began to prepare this Comment, I have
thought a great deal about possible interpretations of (iv).  The authors'
intended meaning is {\em still} not clear to me. 
All interpretations which I have considered are either logically unacceptable
or inconsistent with some part of DJ.

This appendix analyzes the interpretations which I have considered.
I have debated whether it would be worth while to include it,
since I imagine that few readers will be interested in investing their 
time in a detailed  
logical deconstruction of (iv).  I decided to include it for three reasons.

First, it may serve to alert readers to logical problems with 
the statement of the GT even if they choose not to study them in detail. 
Second, I hope that it may motivate the authors of DJ
%
%
to state (iv) precisely in correct logical language in any reply to the 
Comment, so that readers can make informed decisions based on a definite
knowledge of what DJ intended to assume.  Third, since apparently no 
referees' report has yet been received over eight months after submission,%
\footnote{The referee has my sincere sympathy. 
If he is conscientious enough to
try to actually determine the correctness of DJ's densely written proof
based on unclearly stated hypotheses,  
it will take far more time than
is reasonable to ask of an unpaid volunteer. 
}
I hope it may assist the referee.
I assume that the referee who recommended acceptance of DJ 
without clarification of (iv) had not thought carefully 
about its logical meaning.

Recall the hypotheses (iii) and (iv) of DJ's ``General
theorem'' (GT) both as originally posted in arXiv:1106.1871v1 and
subsequently in DJ: 
\begin{quote}
\begin{description}
\item[\rm ``(iii)]  The equality $\dajA = \sum_j \alpha_j (g) \dajEj (g)$
must be satisfied, where the contextual values $\alpha_j (g)$ are 
selected according to the pseudo-inverse prescription.
\item[\rm (iv)] The minimum nonzero order in $g$ for all $\dajEj (g)$ 
is $g^n$ such that (iii) is satisfied.''
\end{description}
\end{quote} 
The statement of hypothesis (iv) is very peculiar, certainly not correct
logical language.  To analyze it, we need to review a few elementary 
principles of logic.

What is the meaning of the statement:
\begin{quote}
\small
``$x = 2$'' \q?
\end{quote} 
I surely hope that the reader mentally replied that it is meaningless
in isolation.  It is a so-called ``open sentence'' to which something
must be added to give it meaning, i.e., to convert it into a logical
statement which is either true or false.  It might also conceivably
be taken as a {\em definition} of the symbol ``$x$'', though this would
not be correct logical language for a definition. 

It can be given meaning by {\em defining} $x$ before stating ``$x=2$''.
For example, if $x$ were previously defined as:
\begin{quote}
\small
Let $x$ denote the largest positive integer which satisfies the equation
$$
x^5 - x^3 - 24 = 0 \q, 
$$
\end{quote}
then ``$x$=2'' would be a logically meaningful statement which would be
definitely true or false (though we might not know which).

To phrase the definition of $x$ just above as  
\begin{quote}
\small
$x$ is the largest positive integer which satisfies the equation
$$
x^5 - x^3 - 24 = 0  
$$
\end{quote}
would not be correct logical language for a definition.  Though some readers
might be able to guess that it was intended as a {\em definition} of 
the symbol $x$, 
it is so far from accepted logical language 
that any logically trained person would
have to question what was meant.  It could not be justified as a logical
shorthand because the previous correctly stated definition is no more
complicated.  This is analogous to (iv) with an inessential change of word
order:  $g^n$ is the minimum nonzero order in  $g  \ \ldots$ . 

In (iv), the symbol $n$ has not been previously defined, so if (iv) is not to
be treated as meaningless, the best guess at the authors' meaning is probably
that (iv) is intended as a {\em definition} of $n$.  But (iv) is supposed
to be a {\em hypothesis} (i.e., a logical statement assumed true), 
not a {\em definition}.

Let us put (iv) aside for the moment to examine another logical principle. 
We noted  that  in isolation, ``$x=2$'' is meaningless as a logical
statement (because there is no way, even in principle, to assign it 
a truth value).  One way to make it meaningful is to predefine $x$.
Another is to prepend one of the so-called logical ``quantifiers'' 
$\forall$ (``for all''
or ``for every'') and $\exists$ (``there exists''), e.g.,
\begin{quote}
$\forall$ integers $x$, $x=2$ (a meaningful statement which happens 
to be false)
\end{quote}
or
\begin{quote}
$\exists$ an integer $x$ such that $x=2$ (a meaningful though somewhat
silly statement which happens to be true).
\end{quote}
In the second statement, the phrase ``such that'' is logically unnecessary 
(and customarily omitted in formal logic), but is added to make the sentence
read well in English.  Also, a necessary specification of 
the so-called ``universe of discourse'' (in
this case that we are talking about integers) has been added to both 
statements.  (If the universe of discourse had been previously specified,
this would be unnecessary.)

Having made these points explicit, let us return to (iv).  
Suppose we temporarily ignore the last clause in (iv) and for purposes 
of examination write the remainder by itself:
\begin{quote}
(iv) The minimum nonzero order in $g$ for all $\dajEj (g)$ 
is $g^n$ such that $\ldots$\q.
\end{quote} 
This is a strange wording which no logician would use, 
so I hate to analyze further without changing the word order 
to obtain more nearly correct
logical language which (so far as I can guess at the authors' intention) 
carries the same logical meaning: 
\begin{quote}
(iv) For all $\dajEj (g)$, the minimum nonzero order in $g$ 
is $g^n$ such that $\ldots$\q.
\end{quote}
For this to begin to make sense, we would have to know what is meant
by ``minimum nonzero order'', which is not a standard
mathematical phrase.   
From the way it is used in the proof of the  GT
and elsewhere,%
\footnote{\lbl{fn10}p.\ 8 just after (5.2) and p.\ 11 just before (7.4).  
Both of
these passages unequivocally imply that given a POVM, one can determine
its ``minimum nonzero order'' $n$ from \re{eq10}, 
without consideration of the "such that" restriction of (iii). 
For example p.\ 11 states  that the ``lowest nonzero order'' for the
counterexample is $n=1$ before considering (iii).  
} 
I assume that it means
that 
\beq
\lbl{eqA20}
\dajEj(g) = \dajE_j^{(0)} + g^n \dajE_j^{(n)} + O(g^{n+1}) 
\q.
\eeq
with the $\dajE^{(m)}_j$ constant operators and $\dajE_j^{(n)} \neq 0$.%
\footnote{A request to the authors to confirm or correct this 
was ignored.}
The statement (iv) just above is still not meaningful because
$n$ remains undefined, but no matter what integer $n$ stands for, 
the statement does imply that {\em all} the $\dajEj (g)$ have
the {\em same} minimum nonzero order $g^n$.  If we take this common minimum
nonzero order as the {\em definition} of $n$, then the statement becomes
meaningful.  Though still strangely worded, it could conceivably be interpreted
as saying that all the $\dajEj(g)$ have the {\em same} minimum nonzero
order, which is to be denoted $n$. 

But what of the restrictive clause beginning ``such that'' 
which we suppressed?  
This clause is 
\begin{quote}
``$\ldots$ such that (iii) is satisfied''.    
\end{quote} 
But we have {\em already} assumed that (iii) is satisfied, 
so the restrictive clause
implies no restriction at all.

%

At this point, an experienced reader will become uneasy, and indeed I did.
I asked myself why the authors would add a restrictive clause
which was no restriction at all.  Could the authors have some other meaning
in mind, but have expressed it in a logically incorrect way?
In trying to guess other meanings, I came up with only one 
plausible possibility, but it turned out to be inconsistent with something
else in DJ.  Next we will examine this possibility.  If a referee did not
think {\em very} carefully about possible meanings for (iv), it might 
superficially seem a reasonable possibility.

One sometimes sees statements in the physics literature 
similar to:
\begin{quote}
\small
$$
e^x = 1+x + x^2/2 \q \mbox{to order $x^2$,}
$$
\end{quote}
meaning that the truncations to order $x^2$ of the Taylor 
series of both sides are equal.
Could (iv) carry a similar meaning?

Well, (iii) is already assumed to hold {\em exactly}, so 
it holds to
all orders in $g$, so if we interpret (iv) as defining $n$ as the smallest
nonzero order to which (iii) holds, then (iv) would define $n:=1$ no matter
what the POVM $\{ \dajEj(g) \}$ was.  It doesn't seem as if that would be the authors' 
intention.  Otherwise, why not simply define $n := 1$?   

Now we enter the realm of real guesswork.  Could (iv) be intended to mean
the following, or something like it?
\begin{quote}
\small 
{\em There exists} a positive integer $n$ such that (iii) holds with each 
$\dajEj(g)$ replaced by its truncation to order $n$, i.e., 
if 
$$\dajEj (g) = \sum_{k=0}^\infty \dajE^{(k)}_j g^k$$ 
and we define $\dajE^\prime_j (g)$ by  
$$
\dajE^\prime_j (g) :=  \sum_{k=0}^n \dajE^{(k)}_j g^k \q,
$$
then (iii) holds with the original $\dajEj (g)$ replaced by 
their truncations $\dajE^\prime_j (g)$,  
$$
\sum_j \alpha_j \dajE^\prime_j (g)  =  \dajA \q, \eqno \mbox{(iii)}^\prime
$$
and moreover, the ``minimum nonzero order $n$'' is 
{\em defined} to be the {\em least} positive integer $n$ for     
which equation (iii){$^\prime$} holds.  
\end{quote}
Note how the quantifier {\em there exists}, which is absent from (iv),
 is very important 
to both the logical  and intuitive meaning.   It alerts the reader that
a nontrivial existence is being assumed, an existence which cannot be taken
for granted.   

Compare the clarity of this with (iv).
Contextual values need not exist.  Even if contextual values do exist 
for a POVM, there is no reason that they should exist for truncations
of that POVM.  (That they do not for the counterexample is basically
DJ's objection to it.)  To assume that contextual values necessarily
exist for some truncations 
is a strong assumption with obscure physical meaning. 
In case (iv) was intended to be interpreted as just described,
how many readers of (iv) would realize that it imposes such a strong  
hypothesis for the ``General theorem''? 

Note also that the definition of ``minimum nonzero order'' proposed by 
\re{eqA20} as a stand-alone phrase has been abandoned.  Now 
\begin{quote} 
\small
``(iv) The minimum nonzero order in $g$ for all $\dajE_j (g)$ is 
$g^n$ such that (iii) is satisfied'' 
\end{quote}
is parsed as 
\begin{quote}
\small
Let ``the minimum nonzero order $n$'' 
denote the least of all positive integers $n$  satisfying [a guess at a 
new meaning of a modified version of (iii) involving $n$], 
where we assume that {\em there exists}
such a positive integer.
\end{quote} 
This newly interpreted (iv) defines ``minimum nonzero order $n$'', 
but the essence of its meaning as a hypothesis 
is the existence assumption.  If this is what the authors of DJ meant, 
then the omission of the quantifier ``there exists'' in (iv) would be a serious
misstatement which would be almost guaranteed to result in misinterpretations.    
All expansions of (iv) from its logical meaning as written to a meaning
which could invalidate the counterexample seem to require some similar
nontrivial  and unstated existence assumption. 

The above is still not logically definite because we have to guess if 
the $\alpha_j$ in (iii)$^\prime$ are the same as already defined 
by the original (iii),
or are defined by the pseudoinverse prescription (which is part of (iii))
applied to the {\em new}, truncated POVM $\{\dajEj^\prime (g)\}$, or 
both.   However, at least it has the potential to give
some interpretation of (iv) in terms of truncations, which is necessary 
for DJ's objection to the counterexample to make sense.

By now the reader's head is probably spinning at the multiplicity of 
conceivable interpretations, but mercifully, we do not have to consider 
all of them in detail.  That is because for the counterexample, there
are {\em no} $\alpha_j (g)$  which satisfy (iii)$^\prime$ for $n=1$,
as DJ shows. 

Therefore, the least $n$ for which (iii)$^\prime$ could hold is $n=2$, 
and it does hold for $n=2$ because the counterexample is quadratic
in $g$
and satisfies (iii).   Therefore, according to the interpretation of (iv)
being considered, the minimum nonzero order $n$ for the counterexample
is $n=2$.

But DJ's objection to the counterexample requires that $n=1$.
I have wondered if DJ may be simultaneously assuming 
two inconsistent
definitions for $n$, the original definition of  \re{eqA20} 
used elsewhere in DJ%
\footnote{see footnote \ref{fn10} 
} 
and the different definition introduced just above.   DJ's objection to the
counterexample is valid only if $n=1$, but the objection also requires
some interpretation of (iv) in terms of truncations.  If (iv) is interpreted
in terms of truncations as above, then $n=2$.  

Of course, I cannot rule out the possibility that DJ may be using some wild
interpretation of (iv) which I haven't even considered.  But in terms of the
above, DJ's objection to the counterexample is invalid. 
Having published ``Sufficient conditions $\ldots$'', unless the authors
withdraw their objection to  the counterexample,  
they have a professional obligation to furnish an unexceptionable
statement of (iv), one which is clear and logically correct.
Only then will readers be able to understand what the GT actually asserts.

All this would become moot if the authors recognize that DJ's attempted proof 
of the GT is in error or incomplete.  But if they come up with a revised 
proof, they should give first priority to restating (iv) 
in a clear and logically correct way. 
\section{Appendix 2:  Misleading stateements in DJ's Section 3} 

This appendix discusses misleading statements in  DJ's Section 3 titled 
``Analysis of a counter-example'' \cite{DJ}.  That section discusses an early 
counterexample \cite{parrott1, parrott4} to a main claim of DAJ \cite{DAJ}.  
Most of what will be said has already been covered in more detail 
in \cite{parrott4}, so  only a brief summary will be given here.
I include it only for completeness, so that DJ's Section 3 will not be 
thought to go unchallenged.  

That 
early counterexample employed three $2 \times 2$ matrices as measurement
operators.  Unlike the later counterexample discussed in the main body
of this paper (which uses three $3 \times 3$ matrices), the earlier 
counterexample does not satisfy the pseudoinverse prescription, which
is explicitly assumed in DJ but not in the earlier DAJ.

My initial interest in DAJ was aroused by the following claim: 
\begin{quote}
{\bf Quote 1 from DAJ:}
``We introduce contextual values as a generalization of the eigenvalues
of an observable that takes into account both the system observable
and a general measurement procedure.  This technique leads to a 
natural definition of a general conditioned average that converges uniquely
to the quantum weak value in the minimal disturbance limit.''
\end{quote}
A counterexample to this claim using three $2 \times 2$ matrices 
was given in \cite{parrott1} and also in \cite{parrott4}.  This counterexample
does not satisfy the pseudoinverse prescription for the simple reason that
there was no reason to think that the authors might have intended it 
intended it as a hypothesis for the claim of Quote 1 above.  
Indeed, I happened to know that they did {\em not} initially intend it as 
a hypothesis because an (incorrect) proof that they sent me 
in response
to an inquiry about DAJ did not assume it; more details are given   
in \cite{parrott4}.

The {\em only} substantive%
\footnote{DJ's claim that ``We devote a considerable amount of space 
to this type of underspecified case [in DAJ] $\ldots$'' is 
at least misleading, and arguably false.
They do devote a long paragraph to a complicated method of calculating
the pseudoinverse of a matrix, but that is standard mathematical material 
which has nothing to do with reasons
for using the pseudoinverse in the first place.  It could have been
replaced by a reference to a standard mathematical text which would have
left space for a genuine discussion of the motivation for the pseudoinverse
prescription. 
}
reference to the pseudoinverse prescription
in DAJ is:
\begin{quote}
{\bf Quote 2 from DAJ:}
\s
``We propose that the physically sensible choice of CV [contextual values]
is the least redundant set uniquely related to the eigenvalues through
the Moore-Penrose pseudoinverse.''%
\footnote{I puzzled for quite a while over the strange language 
``least redundant set uniquely related to the eigenvalues through
the Moore-Penrose pseudoinverse'', which probably carries meaning only
to the authors.  What they mean is \re{eq20} of the present article. 
}
\end{quote}
No reason is given why this should be  ``the physically sensible choice''.
No indication is given that this was to be a hypothesis for the claim
of Quote 1.

DJ suggests that it should be obvious that this is a hypothesis because:
\begin{quote}
[From DJ, not DAJ] ``All the examples we give in the paper [DAJ] use the 
pseudoinverse, and this discussion occurs immediately before the {\em
conditioned average} [italics theirs] section under contention.''
\end{quote}
This is a strange suggestion. 

The ``conditioned average'' section of DAJ has {\em never} been under 
contention.  DAJ's ``conditioned average'' is given by its equation (6),
which does come {\em after} the suggestion of Quote 2 
that the pseudoinverse prescription
is ``the physically sensible choice'', but {\em before} DAJ's 
``Weak values'' section which attempts to justify the claim of Quote 1.
The ``Weak values'' section is what {\em is} under contention.  

The short proof  of DAJ's  expression (6) giving its 
``conditioned average''  does 
{\em not} require the hypothesis that the contextual values satisfy the
pseudoinverse  prescription.   Do the authors contend that readers should
conclude that the pseudoinverse prescription (PIP) was intended as a hypothesis 
for (6) just because Quote 2 came before (6) and even though the PIP is 
unnecessary for its proof?

The first arXiv version of DJ (which 
differs substantially from the published version), 
arXiv:1106.1871v1, gives the first
indication 
of why the authors consider the pseudoinverse prescription 
``the physically sensible choice''.  
It is because the pseudoinverse
prescription minimizes a particular {\em upper bound} 
for the detector variance.  It does not minimize the detector variance 
itself, and the upper bound which it does minimize is not the best possible
upper bound.  This situation is discussed in detail in \cite{parrott4},
including a derivation of a sharper upper bound, and the discussion will
not be repeated here. 

It is worth emphasizing that DAJ gives {\em no reason whatever} to 
 use the pseudoinverse prescription (PIP), and the {\em only}
 reason that DJ gives  
is that it minimizes a particular upper bound for the detector variance.
But this has absolutely nothing to do with weak values.
It is hard to see how the authors could imagine 
that readers of DAJ should obviously realize that the PIP should be intended
as a hypothesis for the ``Weak values'' section and the claim of Quote 1. 

Call the upper bound which the pseudoinverse prescription
does minimize the ``DJ upper bound'' (DJUB), to distinguish it from other possible
upper bounds such as the better upper bound of \cite{parrott4}.
Section 3 of
DJ explicitly carries out a messy calculation of 
the DJUB for both the contextual values of the $2 \times 2$ 
counterexample and also for those determined by the pseudo-inverse 
prescription, and finds that the latter is smaller.  
But this is built into the abstract mathematics of the Moore-Penrose 
pseudoinverse; 
there was no need to calculate it explicitly.  

DJ goes on to intimate that the pseudoinverse contextual values 
 are  physically
superior to the contextual values of the counterexample, even though this
does not follow from the fact that the particular DJUB upper bound for 
the pseudoinverse prescription is no more than 
the DJUB for any other choice of contextual values
(including that of the counterexample).%
\footnote{Of course, the fact that an upper bound for quantity A 
is less than an upper bound for quantity B does not imply that 
quantity A is less than quantity B.
}
However, even if it {\em did} follow it would be
{\em completely irrelevant} to the counterexample.   
The $2 \times 2$ counterexample was 
devised as a counterexample to the Quote 1 mathematical claim of DAJ,
that its ``general conditioned average $\ldots$ converges uniquely'' to
the traditional weak value.  The counterexample was posted well before 
DJ revealed why
the authors consider the pseudoinverse
solution ``the physically sensible choice''.  
The counterexample was never claimed to have 
any physically desirable properties.


Section 6.3 of \cite{parrott4} introduces a simple method 
of determining contextual values which 
minimizes the detector variance itself, not just DJ's particular [DJUB] 
upper bound
to the detector variance.  It does require knowing the probabilities of the 
various detector outcomes, but this information is always available to 
an arbitrary approximation (otherwise there would be no way to use the 
detector to approximate the expectation of the system observable).  If the 
authors want to advocate for the pseudoinverse prescription, it would have
been more appropriate to compare it to the just-mentioned suggestion of 
\cite{parrott4} than to a counterexample which was devised for a completely 
different purpose.  

\section{Afterword}

Since I first became interested in the claims of DAJ over a year ago, 
pursuit of those claims has turned into something of a nightmare.  
The present paper is sufficiently technical that I imagine that it will have 
few, if any readers.  Yet it seems that it had to be written.

I would be surprised if the various papers by the DJ authors 
have many readers, either.  I have
{\em never} encountered any reader who has demonstrated knowledge of DAJ or DJ 
sufficiently detailed to discuss it in depth.  

I have absolutely no doubt that any controversy over DJ's 
``General theorem'' (GT)
could be quickly settled by discussions between any two people who had 
actually scrutinized its attempted proof.  If the authors 
would discuss it with me, we could probably settle the matter in very little
time, and that it how it 
should be settled.  Only in the unlikely event that some fundamental 
disagreement remained would it be necessary to involve journals and referees.
It seems wrong that the time of so many people is being unnecessarily wasted.

The physics literature is full of papers with questionable claims which 
were probably routinely published with little or no meaningful scrutiny.
Unless of obviously fundamental importance, 
such claims are rarely withdrawn or corrected.  It is easy to get such
papers published if the authors have appropriate credentials, but very 
difficult to question them in print.  
Meaningful ``peer review'' is minimal.

DJ may turn out to be yet another example.  (I could furnish several.)  
DJ was accepted almost immediately, 
within a month of submission,
but the present Comment has been in limbo for almost nine months, with 
no indication why. 
That does not augur well for eventual acceptance.

I have no definite reason to think that it will be rejected, but I will
not be surprised if it happens.  I {\em would} be surprised if the rejection
were for substantive reasons, such as a demonstrable error.  Based on 
experiences with other journals, I would instead expect the stated reason
to be some vague value judgment which would be impossible to contest, 
such as ``too mathematical'' or ``too long for a `Comment' '' or 
``uninteresting''.   

Indeed, a journal may not even bother with a pretext.    
I have had the experience of a journal (not J.\ Phys.\ A) keeping a paper 
for over a year, with
uncommunicative responses to inquiries every few months about its status,
only to finally reject it with no referee's report, no reason whatever 
given, no indication that it had ever been read, and no apology (not 
for rejecting it, but for the way it was handled).  
A subsequent inquiry as to the reason for the rejection went unanswered.
\section{Appendix 3:  The ``Corrigendum''\protect{\cite{corrigendum}} 
to DJ} 

\subsection{Introduction} 
\noindent
This appendix was added around February 18, 2013. 
\\[.5ex]

Over a year after the present work questioned the proof of the GT, 
the authors added a December, 2012, ``Corrigendum'' 
\cite{corrigendum}
acknowledging a gap in its attempted proof. 
The sole content of the Corrigendum is a Lemma which, in the context of
DJ's attempted proof of its GT, is essentially equivalent to the Conjecture
which is the centerpiece of the present work.  
The Corrigendum does not mention the Conjecture.  
%

This appendix points out a simple calculational error in the Corrigendum
which invalidates its proof of its Lemma.  
(I think that there is a more subtle error as well.) 

Then it will present an 
elementary proof of an important special case of the Conjecture/Lemma.
The special case may include most situations  
of physical interest.  

%
\subsection{Review of notation and definitions}
\lbl{subsec6.2}
The Corrigendum's Lemma assumes a POVM $\{ \dajEj (g) \}$ whose elements
are {\em linear} polynomials in the ``weak measurement'' parameter $g$:
\beq
\lbl{A3eq10}
\dajEj(g) = \dajE^{(0)}_j + g\dajE^{(1)}_j \mbox{where 
$\dajE^{0}_j$ and $\dajE^{(1)}$ are constant operators.}
\eeq
(Actually, it replaces $g$ with $g^n$ for some positive integer $n$,
but this is obviously equivalent and only complicates the notation.)

Recall that the ``contextual values'' $\alpha_j(g)$ are chosen 
to satisfy
\beq
\lbl{A3eq20}
\dajA = \sum_j \alpha_j (g) \dajEj (g)
\q,
\eeq 
where $\dajA$ is the ``system observable'' whose expectation in a given
state is to be weakly measured.  
This equation will not always have a solution, but DJ considers only
situations in which it does.  When there are multiple solutions, DJ
requires that 
the so-called ``pseudoinverse prescription'' be applied to single out 
a unique solution. 

It is also assumed that everything in sight is finite dimensional 
and that the POVM elements $\dajEj (g) $ and the 
system observable $\dajA$ mutually commute, so that they are represented
by finite, diagonal $m \times m $ matrices:  
$\dajEj (g) = \diag 
(E^1_j(g), \ldots , E^m_j (g) )$, 
$\dajA = \diag (a_1, \ldots , a_m)$.  
Also write $\vec{E_j} := 
(E^1_j(g), \ldots , E^m_j (g) )^T$, 
and 
$\vec{a} := (a_1, \ldots , a_m)^T.$  
In terms of this notation,
equation \re{A3eq20} can be written
\beq
\lbl{A3eq30}
F(g) \vec{\alpha}(g) = \vec{a} \q,
\eeq   
where $\vec{\alpha}(g) := (\alpha_1(g), \ldots, \alpha_m(g))^T,$ and   
$F = F(g) := [\vec{E}_1 (g), \ldots , \vec{E}_m (g) ]$ denotes the 
matrix whose columns are the $\vec{E}_j$.  Moreover, $F$ is a linear
polynomial in $g$:
\beq
\lbl{A3eq40}
F (g) =  P + g R 
\q,
\eeq
where $P$ and $R$ are constant matrices.

The Corrigendum's notation replaces $R$ with $F_n$ (and also $g$ with 
$g^n$ as previously noted), but this seems to invite confusion. 
I am trying to stay as close as possible to the Corrigendum's notation,
but I find it hard to bring myself to add gratuitous complications 
such as replacing
$g$ by $g^n$ and using the potentially confusing $F_n$ for 
the linear homogeneous part of $F$.
Nevertheless, the subsection analyzing the Corrigendum's
proof will use the Corrigendum's notation exclusively to avoid confusion.



%
\subsection{Relation between our Conjecture 
and the Corrigendum's Lemma}

DJ defines the {\em singular values} of a matrix $F$ as the
eigenvalues of $(FF^T)^{1/2}$.  
We shall be interested in the case in which
$F = F(g)$ is as defined as above to be the matrix which determines the
contextual values via the equation 
$F(g) \vec{\alpha} (g) = \vec{a}$. 

One form of the 
polar decomposition of the real matrix $F(g)$ is 
\beq
\lbl{A3eq50}
F(g) = (F(g)F(g)^T)^{1/2} V(g) 
\q,
\eeq
with $V(g)$ a partial isometry, with initial space the initial space 
In($F(g)$) of
$F(g)$ and final space the range of $F(g)$.%
\footnote{
The {\em initial space} In($F$) of a matrix is the orthogonal complement
of its nullspace, which is the subspace spanned by its rows. 
A {\em partial isometry} is an operator which is an isometry when
restricted to its initial space.  Its range is called its {\em final space}.
\s
An {\em isometry} is an operator which preserves
norms (and, consequently, inner products), but is not necessarily
surjective.  A surjective isometry on a real (complex) Hilbert space is called 
{\em orthogonal (unitary)}.
In the typical case in which the $F(g)$ are invertible, $V(g)$
is orthogonal. 
} 
If the nullspace of $F$ is nontrivial, of course $F^{-1}$ cannot 
exist in the strict sense, 
but we can define a (Moore-Penrose) ``pseudoinverse'', temporarily 
also denoted $F^{-1}$, as the operator which acts on the range of $F$
as the inverse of  
$F |\, \mbox{In}(F) : \mbox{In}(F) \rightarrow \mbox{Range}(F)$,
and annihilates the orthogonal complement of the range of $F$. 
In other words, for $ w \in \mbox{In($F$)},  F^{-1}F (w) := w  $, 
and for $v$ orthogonal to Range($F$), $F^{-1}v := 0$. 

Then
\beq
\lbl{A3eq60}
F(g)^{-1} = V(g)^{-1}(F(g) (F(g)^T)^{1/2})^{-1} 
= V(g)^T (F(g)F(g)^T)^{-1/2}
\q.
\eeq
Since $V(g)$ is bounded for all $g$,
the contextual values 
\beq
\lbl{A3eq70}
\vec{\alpha} (g) := F(g)^{-1} \vec{a}
\eeq
(which will exist iff $\vec{a}$ is in  Range($F(g)$) = In($F(g)F(g)^T$) 
for small $g$, a basic assumption of DJ) 
will be $O(1/g)$ if (and, because of the special form of $V(g)$, only if) 
the nonzero singular values of $F(g)$ (assumed
analytic in $g$) have leading order no greater than $g^1$.   
The conclusion of the Corrigendum's Lemma is that the nonzero singular values 
do have leading order no greater than $g^1$.  Then 
as noted in Section 2.2.3, $\vec{\alpha}(g) = O(1/g)$ implies the Conjecture.%
\footnote{This fact may not be easy to extract from published literature.
It is the essence of DJ's attempted proof of the GT, but that is not 
easy to penetrate.  It also follows from equation (113) of  \cite{parrott4}. 
} 


I have reason to believe,  but have not checked,
that analyticity of the singular values follows from nontrivial theorems
concerning analytic perturbations of Hermitian matrices.  (The Corrigendum
does not address the analyticity, seeming to take it for granted.) 
I have not checked it because relevant references (e.g., a 1969 book
of Rellich) are not readily available,%
\footnote{I live 200 miles from the nearest research library.}
and 
my proof of the Conjecture 
established by means not requiring analyticity 
that the contextual values
are $O(1/g)$.  
Because of that, I consider, 
the Lemma and Conjecture as 
essentially equivalent in the context of the proof of the GT.%

For the reader's convenience, we reproduce the Corrigendum's statement 
of the Lemma
and the first few lines of its proof.
\\[2ex]
\begin{quote}
\noindent 
``{\bf Lemma.} {\em
The singular values of the $M \times N$ dimensional matrix 
$F = P + g^n F_n$ with $M \leq N$ have maximum leading order
of $g^n$, where $P = [p_1 \vec{1} \ldots p_n \vec{1} ]$ and
$F_n = [ \vec{E}_1 \ldots \vec{E}_N ]$ such that
$\sum_j p_j = 1$ and $\sum_j \vec{E}_j = \vec{0}.$
} 
\\[1ex]
{\bf Proof.}  \ldots . Since $P^T F_n = 0$, the latter has the 
simple form $H = P^T P + g^{2n} F^T_n F_n$, \ldots '' 
[The Proof had previously defined $H := F^T F$.]
\end{quote}
The authors don't further explain their notation, but it is clear from
the context that $\vec{1}$ stands for the column vector whose entries
are all 1, and  
$[ \vec{E}_1 \ldots \vec{E}_N ]$ for the matrix whose columns are 
the vectors $\vec{E}_j$.
An example of such  $P$ and $F_n$ is: 
$$
P := 
\left[ 
\begin{array}{cc}
1/2 & 1/2 \\
1/2 & 1/2 
\end{array}
\right], \q 
F_n := 
\left[ 
\begin{array}{cc}
2 & -2 \\
2  & -2 
\end{array}
\right]
\q. 
$$
Then 
$$
P^T F_n  = 
\left[ 
\begin{array}{cc}
2 & -2 \\
2  & -2 
\end{array}
\right] = F_n \neq 0. 
$$
The Corrigendum's proof does not actually require that $P^T F_n = 0$,
but only that $P^T F_n + F^T_nP = 0$.  However, that is also false:
$$
P^T F_n + F^T_n P = F_n  + F^T_n = 
\left[ 
\begin{array}{cc}
4 & 0 \\
0  & 4 
\end{array}
\right] \neq 0. 
$$

The proof's assertion that ``$P^T F_n = 0$'' {\em would} be true if the
{\em columns} (instead of just the rows) of $F_n$ also summed to 0 .  
I confess that
  in my first readings of the proof, I succumbed to that confusion 
in mental calculations of the matrices, and 
believed that the subsequent proof could be correct.
But the claimed form of $H = P^T P + g^{2n} F^T_n F_n$ seems essential
to the subsequent analysis, so I think the proof has to be considered 
invalid, or at least invalid {\em as written}.  
\subsection{Elementary proof of a special case of the Lemma/Conjecture}

This subsection presents a straightforward elementary proof of what may be 
the most important special case of the Lemma, namely the case in which
$F = F(g)$ is an $N \times N$ (i.e., square) matrix of the 
``linear'' form \re{A3eq40} and is  invertible 
for $g$ near $0$.  When $F$ is square, which is probably the typical
case for applications, invertibility near zero is the ``generic'' case.

We shall show that for square, invertible $F(g)$, 
the contextual values are $O(1/g)$.  
We shall do this using Cramer's rule to obtain the 
contextual values.


We shall observe that for an arbitrary ``linear'' $F = F(g)$ (not necessarily
invertible) which arises from a POVM,
 $\det F(g)$ is proportional to $1/g^{N-1}$.  
(Of course, the constant of proportionality is nonvanishing if and only if 
$F(g)$ is invertible.) 


By definition, a ``linear'' $F= F(g)$ is of the form
\beq
\lbl{A3eq90}
F(g) = P + gR
\q,
\eeq
where $P$ and $R$ are constant matrices. 
In addition, an $F(g)$ arising from a POVM as described 
in subsection \ref{subsec6.2} 
satisfies the following conditions.
\beq
\lbl{A3eq95}
P =
\left[
\begin{array}{cccc}
p_1 & p_2 & \ldots & p_m \\
p_1 & p_2 & \ldots & p_m \\
\     & \     & \ldots & \     \\
p_1 & p_2 & \ldots  & p_m
\end{array}
\right]
\q,
\eeq
where $\sum_j  p_j = 1$, in order to assure that
$\{ \dajE_j \}$ satisfies the defining condition $\sum_j \dajEj = \dajI$ for
a POVM, with $\dajI$ the identity matrix.  
That all columns of $P$
are multiples of $(1, 1, \ldots , 1)^T $  follows from DJ's hypothesis
(i) for its ``General theorem'': that all measurement operators be nearly
proportional to the identity for small $g$  to ensure that the measurement
be weak for small $g$.  Similarly, the POVM condition implies that
all rows of $R$ sum to zero: $\sum_j R_{ij} = 0$ for all $i$.  

For an $N \times N$ matrix $M$, 
let $\vec{M}_j$ denote its $j$'th column, and write
$M = [\vec{M}_1 \ldots  \vec{M}_N]$. 
To compute $\det F(g)$, we think of the determinant of a matrix 
as a multilinear function of its columns.  Using multilinearity to 
expand the determinant, gives it as a sum of terms of the form 
either $\det P$ or $g^N \det R$ or
\beq
\lbl{A3eq100}
\pm g^{N-k} \det [\vec{P}_{\pi(1)}, \vec{P}_{\pi(2)}, \ldots , 
\vec{P}_{\pi(k)},
\vec{R}_{\pi(k+1)}, \ldots , \vec{R}_{\pi(N)} ] 
\q,
\eeq
where $\pi(1), \ldots , \pi(N)$ is a permutation of $1, \ldots , N$
and $1 \leq k \leq N-1$.

Because $P$ has rank 1 (columns proportional), 
 $\det P = 0$, and because the rows of $R$ all sum to 0 (so that its
columns are linearly dependent), also $\det R = 0.$ 
Also because $P$ is rank 1, 
the only nonvanishing
terms of \re{A3eq100} 
contain just one column of $P$, i.e., $k=1$.  
This shows that $\det F(g)$ is proportional to $g^{N-1}$.

For future use, let us revisit the argument just given without using the 
assumption that the rows of $R$ sum to zero.  In this case,
we need not have $\det R = 0$, so we would conclude that 
$\det F(g) = (a + bg)g^{N-1} $ with $a, b $ constant.

All $(N-1) \times (N-1)$ ``minor matrices''%
\footnote{A ``minor matrix'' $M^{(ij)}$ of $F$ is the 
$(N-1) \times (N-1) $ matrix obtained from $F$ by 
deleting the $i$'th row and $j$`th column.
}
$M^{(ij)} (g)$ of $F(g)$ 
are of the ``linear'' form \re{A3eq90}:
\beq
\lbl{A3eq110} 
M_{(ij)}(g) = P^{(ij)} +  g R^{(ij)}
\q,
\eeq 
but the $(N-1) \times (N-1)$ matrices
$P^{(ij)} $ and $R^{(ij)}$ need not satisfy all of the auxiliary conditions
of the original $P$ and $R$.  However, the $P^{(ij)}$
are still rank 1, so that their columns are proportional.
In general, the rows of the $R^{(ij)}$ need not sum to zero.

Cramer's rule gives the  contextual values 
$\alpha_j (g) = (F^{-1} (g) \vec{a})_j$ as a sum of terms, each of which
is of the form
$$
\alpha_j (g) = (F^{-1} (g) \vec{a})_j = \frac{\pm a_k
\det M^{(jk)} }{\det F(g)}
\q
$$ 
Since we have already established that  $\det F(g)$ is 
proportional to $g^{N-1}$, 
to show that $\alpha_j (g) = O(1/g)$ it suffices
to show that 
\beq
\lbl{A3eq120}
\det M^{(jk)} = g^{N-2} (c_{jk} + O(g)) \q 
\mbox{with $c_{jk}$ constants (possibly zero).}  
\eeq
Evaluating the minor matrices as we evaluated $\det F(g)$ gives a sum 
of terms, each of which is proportional to either $g^{N-1-1} = g^{N-2}$
as before plus one more term equal to $g^{N-1} \det R^{ij}$.  
(This additional term
arises because the rows of $R^{(ij)}$ need not sum to zero, so 
we cannot conclude that $\det R^{(ij)} =  0 $ as in the calculation
of $\det F(g)$.) 
Each of these terms is of the desired form \re{A3eq120}. 

An earlier section mentioned that I believed I had proved the 
full Conjecture (not just the special case above): 
\begin{quote}
\small
``I have sketched such a proof [of the Conjecture]  but have not
written it in detail, so I make no claims.  I will be happy to share 
the ideas of the proof with any qualified person who might be interested 
in expanding on them.  They are not difficult, but
annoyingly detailed.''  
\end{quote}
The annoying details arise when one tries to generalize to cases 
when $F(g)$ may not be invertible, or even square.  Here one has to 
use DJ's hypothesis that the contextual values satisfy the pseudoinverse
prescription.  I hesitated to write up that part because 
the authors' stated reason for using
the pseudoinverse prescription is so unconvincing, as explained 
in Appendix 2 and \cite{parrott4}.  
So far as I can see, the only reason
to consider it would be to obtain the traditional weak value for 
linear POVM's.  
I am  not sure why that would be important,
so I decided to invest my time otherwise.

\subsection{What if the authors succeed in proving the Corrigendum's Lemma?}

I came across the Corrigendum by accident only a few days ago---the
Journal of Physics A never notified me, much less asked my opinion of it. 
This appendix was originally intended to discuss the proof of its Lemma,
which seemed to me not quite correct, though I thought I knew how to correct
it.  I was already somewhat familiar with the Lemma via its previous
appearance in \cite{DJ2}, and believed it to be essentially correct.
I had not noticed the simple error exposed above.  

In the course of writing the Appendix, I did notice the simple error.
It is not the sort of error that I would expect from the authors
of the Corrigendum, and as of this writing, I find it hard to believe.
I keep wondering if {\em I} might be committing some simple error,
perhaps an overlooked hypothesis or something similar that would be 
hard for an author to notice. 

Moreover, the error does not show that the 
{\em conclusion} of the Lemma is wrong, only 
that its proof {\em as written} is incorrect.    
It is conceivable that the authors could later
produce a correct proof.  

Even if so,
I would still have to question the Corrigendum.  It does not address
the proof gap discussed at the end of Section 2.2 of Version 6.  Even  
a correct Lemma would prove no more than the Conjecture which 
is the centerpiece of the present work---that the conclusion of the GT
holds for {\em linear} POVM's.   It would not prove the full GT claimed 
by DJ. 

The Corrigendum closes with an ``Acknowledgement'' thanking Version 6
of the present work for 
``urging us to justify this lemma''.  Never publicly or privately
did I urge the authors ``to justify this lemma''.  
My submitted ``Comment'' could be taken as urging them to justify 
the proof of the GT, but that is not the same as justifying the Lemma. 

I find it 
troubling that many might interpret  the Acknowledgement 
as claiming my implicit
endorsement for the truth of the GT. 
Do the authors not realize that ``this lemma'' does {\em not} establish
the full GT, but only the Conjecture?  Or do they believe that the Lemma
{\em does} justify the full GT, but that Version 6's objections to 
the GT's proof (at the end of Section 2.2) are unworthy of reply?

%


\begin{thebibliography}{0}
\bibitem{parrott1}
Parrott, S.\ `Contextual weak values' of quantum measurements with
positive measurement operators are not limited to the traditional
weak value, arXiv1102.4407
\end{thebibliography}
\end{document}